# The Upper Critical Field $H_{c2}$ in Advanced Superconductors with Anisotropic Energy Spectrum

*M. E. Palistrant*
*Institute of Applied Physics, Academy of Sciences of Moldova, Chisinau, 2028 Moldova*
*\*e-mail: mepalistrant@yandex.ru*

*Abstract*

A brief review of works on the microscopic theory of determining the upper critical field in two-band isotropic and anisotropic superconductors is given. The research is based on a set of the Ginzburg-Landau equations for the order parameters in a magnetic field that are studied in terms of the classical approach to a superconducting system in a magnetic field. Two inequivalent energy bands with different topology of Fermi surface cavities overlapping on the Fermi surface are discussed. The cases of the direction of the external magnetic field $\vec{H}$// the (ab) plane and $\vec{H}$ // the crystallographic *c* axis are studied. The equations for determining $H_{c2}$(ab) and $H_{c2}$(c) for a pure superconductor and a superconductor doped with electrons and holes are derived. The analytical solutions to these equations in the vicinity of the superconducting transition temperature ($T_c$ – T<<$T_c$) and in the vicinity of zero (T<<$T_c$) are found. The temperature and impurity dependences of the upper critical fields $H_{c2}$(ab) and $H_{c2}$(c), as well as the anisotropy coefficient $\gamma_H$, are studied. The resulting theory is applied to determine the dependences of the above magnetic characteristics of intermetallic compound $MgB_2$. The theory agrees qualitatively with experimental data.

## 1. Introduction

The discovery of the mechanism of superconductivity [1] was followed by a rapid development of the theory of this amazing phenomenon. Numerous theoretical studies have resulted in the so-called BCS–Bogoliubov-Gor'kov-Eliashberg theory. This list may comprise also names of other eminent scientists. The above studies were carried out on the basis of isotropic models and were intended for explaining the properties of superconducting metals and alloys.

In addition, the study of more complex systems required to take into account the effect of different kinds of anisotropy on their superconducting properties, in particular, the anisotropy of the electron energy spectrum, i.e., the presence of the overlapping of two and more energy bands on the Fermi surface (multiband case). In recent years, the difference in the topology of Fermi surface cavities of the bands under study was also taken into account. The two-band model was propozed in [2] and idepenly in [3]. The main assumption of the model is the formation of the Cooper pairs of electrons inside one energy band and transition of this pair as a whole entity from one energy band to another. This results in the appearance of the intraband $V_{nm}$ (m≠n, n;m=1,2) electronic interactions that lead to the additional attraction of electrons which, in its turn, favors an increase in the superconducting transition temperature. This is equivalent to the diagonal approximation over band indices in the basic equation of the theory; two order parameters $\Delta_1$ and $\Delta_2$ appear in the two-band model.

Having made these assumptions, Moskalenko and his associates carried out investigations of the thermodynamic and electromagnetic properties of many-band superconductors. A few books and a lot of articles on this problem were published.
Let us mention some books [4-7], reviews [8-10], and articles [11-14] concerned with the history of multiband superconductivity. These works, and especially [11-14], list the names of scientists from various countries who have made their contributions to the development of the theory of two-band superconductivity. In particular, we should mention profound reviews [15-17].

The study of the overlapping of energy bands yields not only the quantitative difference of the results from the case of a one-band superconductor, but, in some cases, qualitatively new results. For example:

(1) In two-band superconductors, high temperatures of the superconducting transition can take place not only in the case of the attractive interaction between electrons but even in the case of repulsive one ($\lambda_{nm} < 0$, n; m = 1-2), yet the relationship $\lambda_{11}\lambda_{22} - \lambda_{12}\lambda_{21} < 0$ is fulfilled [18].

(2) In impurity two-band superconductors, for example, the Anderson theorem is violated at $\Delta_1 \neq \Delta_2$, and the dependence of the thermodynamic quantities on the concentration of a non-magnetic impurity due to the interband scattering of electrons on impurity atoms is observed [4,19].

(3) In two-band superconductors, collective oscillations of the exciton-like Leggett mode caused by phase fluctuations of order parameters from different bands are observed. The number of such oscillatory modes can be two in three-band systems as well as in a two-band system that is reduced to an effective three-band model with regard for the Cooper pairs of electrons from different energy bands [20, 21].

(4) Using the two-band model and assuming moderate values of the coupling constants, one can obtain high values of $T_C$, two energy gaps $2\Delta_1/T_C > 3.5$ and $2\Delta_2/T_C < 3.5$, large values of negative $dlnT_C/dlnV$ (V is the volume), a positive curvature of the upper critical field near the transition temperature, etc. [9]. Furthermore, in the two-band model, it is possible to describe the decrease in $T_C$ with increasing system disorder [22].

(5) The position of the Fermi level, which can be changed by doping, plays an important role in the determination of the thermodynamic and magnetic properties of a two-band superconductor. Having assumed the non-phonon pairing mechanism of superconductivity, as well as the phonon mechanism in the many-band systems with lowered densities of charge carriers, the regard for the singularities mentioned above is extremely crucial. A particular interest is given to the possibility of obtaining a bell-shaped dependence of $T_C$ and the jump of heat capacity $(C_S - C_N)/C_N$ at the point $T = T_C$ on the carrier density [7]. In the three-band model, it is possible to obtain a "step" in the dependence of $T_C$ on the carrier density.

An expansion in the number of energy bands on the Fermi surface increases the overall density of electron states and leads to the onset of an additional interband electron-electron interaction that contributes to the onset of superconductivity. This interaction violates the universal BCS relations and leads to the substantial dependence of some physical characteristics on the properties of an anisotropic system.

All of the above classical results had been obtained long before the discovery of high temperature superconductivity in $MgB_2$ [23].

This work is concerned with the determination of one of the important characteristics of the superconducting state, i.e., the upper critical field $H_{c2}$ in both isotropic and anisotropic two-band systems. We call a two-band system isotropic if the Fermi surface cavities of the two bands are spherical and we call it anisotropic if they differ, for example, due to different dimensions of overlapping energy bands.

In this work, we will construct a microscopic theory of determining the upper critical field for both isotropic and anisotropic two-band systems. The need to solve this problem arose after the discovery of high temperature superconductivity in compound $MgB_2$ that exhibits significant anisotropy of the upper critical field $H_{c2}$ (see, e.g., [24]).

It should be noted that the upper critical field $H_{c2}$ was calculated by Gor'kov [25] using the microscopic theory of superconductivity for a pure one-band superconductor at temperatures close to $T_c$. Subsequently, this theory was generalized by Maki and Tsuzuki [23] for one-band systems to the entire temperature range $0 < T < T_c$.

The development of the two-band superconductivity theory (after publications [2, 3]) also necessitated the construction of a microscopic theory for the upper critical field $H_{c2}$ for a pure two-band superconductor. Such a theory had been proposed for the first time by Palistrant and Dedyu [27] long before the discovery of superconductivity in $MgB_2$.

Thus, the anisotropy of the system requires the construction of a new theory (in particular, for the magnetic properties of intermetallic compound $MgB_2$).

We are aware of two theoretical publications in which the behavior of $H_{c2}$ is described for a pure two-band superconductor with anisotropic properties similar to $MgB_2$.

It is referred to [28] and [29]. In either work, an original approach and technique are used. For example, the value of $H_{c2}$ in [28] is determined using the multiband formulation of the Elenberg semiclassical theory [30]. Owing to strong anisotropy, the values of $H_{c2}^{(ab)}$ and $H_{c2}^{(c)}$ differ considerably, which makes it possible to obtain results for pure $MgB_2$ close to experiment. It was shown in [29] that ratio $H_{c2}^{(ab)}/H_{c2}^{(c)}$ increases upon cooling.

The aim of this paper is to construct a microscopic theory of determining the upper critical field $H_{c2}$ on the basis of the fundamental principles of the theory of superconductivity for a system in a magnetic field [25, 26].

Below, we focus our attention mainly on the overlapping of energy bands on the Fermi surface (multiband case) and on anisotropy due to different dimensions of the energy bands under study. This circumstance, in its turn, leads to different topologies of the Fermi surface cavities.

The studies below are based on a set of Ginzburg-Landau equations for the order parameters of a two-band system in a magnetic field [31, 32]; the techniques of calculating the $H_{c2}$ values for isotropic and anisotropic two-band systems are developed and generalized. The equations defining $H_{c2}$ and $T_c$ are derived, the asymptotic solutions for the value of $H_{c2}$ are found; the $H_{c2}(T)$ curves are studied in the entire temperature range $0 < T < T_c$ both in isotropic and anisotropic two-band systems.

Note that the regard for only one type of anisotropy (the overlapping of energy bands on the Fermi surface) [27] leads to an important, qualitatively new result in comparison with a single-band system; that is, in the $H_{c2}$ dependence on T, a positive curvature appears in the vicinity of the superconducting transition temperature. This phenomenon results from different values of the average velocity of electrons in the *n*th cavity of the Fermi surface. Another anisotropy due to the inequivalence of the energy bands under study leads to a significant difference of magnetic field $H_{c2}(ab)$ from $H_{c2}(c)$. In addition, the value of $H_{c2}(ab)$ is several times greater than the value of $H_{c2}(c)$. A significant role in this problem is played by the two-dimensionality of the σ-band of intermetallic compound $MgB_2$ and small values of the average velocities of electrons on the Fermi surface in the direction of the *z* axis. In this work, we analyze both the case of isotropic two-band systems and anisotropic ones with the direction of the magnetic field H ∥ (ab) and H ∥ (c). The temperature dependence of the anisotropy coefficient is revealed, and a qualitative agreement of the results with experimental data is shown. We have found that it is possible to determine the effect of the mechanism of filling of energy bands on the physical characteristics of an $MgB_2$ system upon its doping with electrons or holes.

In this work, the results are presented briefly and concisely, they allow understanding the overall picture of occurring processes. For more details of this problem, in addition to [9], see [27] and [33-36].

## 2. Determination of the upper critical field $H_{c2}$ in an isotropic two-band system

The study is based on a set of the Ginzburg-Landau equations for the order parameters $\Delta_m(\vec{x})$ in a magnetic field [31].

$$\Delta_m^*(\vec{x}) = \frac{1}{\beta}\sum_{nn'} V_{mn} \int d\vec{y} \sum_\omega g_{n'n}(\vec{y},\vec{x},\omega)\Delta_{n'}^*(\vec{y}) g_{n'n}(\vec{y},\vec{x},-\omega) -$$
$$-\frac{1}{\beta}\sum_\omega \sum_{n,n'}\sum_{p,p'} V_{mn} \int d\vec{y} \int d\vec{y'} d\vec{y''} \Delta_n^*(\vec{y}) g_{n'n}(\vec{y},\vec{x},\omega) \Delta_p(\vec{y'}) \times$$
$$\times g_{n'p}(\vec{y},\vec{y'},-\omega)\Delta_{p'}^*(\vec{y''}) g_{\vec{p'}\vec{p}}(\vec{y''},\vec{y'},\omega) g_{p'n}(\vec{y''},\vec{x},-\omega). \qquad (2.1)$$

Here (m; n = 1.2), $V_{mn}$ are the effective matrix elements of the electron-electron interaction. The Green function $g_{n'n}$ in the presence of the magnetic field $\vec{H}$ is defined by the equation [25, 26]

$$g_{n'n}(\vec{r},\vec{r'},\omega) = e^{i\varphi(r,r')} g_{n'n}^0(\vec{r},\vec{r'},\omega), \qquad \varphi(r,r') = e\int_{r'}^{r} A(\vec{l})d\vec{l}. \qquad (2.2)$$

where $g_{n'n}^0$ is the Green function of an electron in a normal metal without magnetic field. The presence of a magnetic field is taken into account by the phase multiplier.

In equation (2.1), we expand the normal metal function $g_{n'n}^0$ into a series by the Bloch function $\psi_{n\vec{k}}$:

$$g_{n'n}^0(\vec{y},\vec{x},\omega) = \sum_{\overline{k}\overline{k'}} g_{n'n}^0(\vec{k'},\vec{k},\omega) \psi_{n'\vec{k'}}(\vec{y})\psi_{n\vec{k}}^*(\vec{x}). \qquad (2.3)$$

The nonzero solutions in equation (2.1) correspond to the origination of the superconducting state. At the point H = $H_{c2}$, solutions with infinitely small quantities of the order parameter $\Delta_m^*$ appear for the first time, which makes it possible to significantly simplify this set of equations being restricted to the linear terms $\Delta_m^*$.

We shall use the approximation of diagonal Green functions (see, e.g. [4, 31]) and choose the vector potential $A(\vec{r})$ in the form $A_x = A_z = 0$; $A_y = H_0$ x (the magnetic field is directed along the *z* axis). Equation (2.1) takes the following form

$$\Delta_m^*(\vec{r}) = \sum_n V_{nm} \frac{1}{\beta}\sum_\omega \int d^3(r')\Delta_n^*(\vec{r'}) \int \frac{d^3k\, d^3q}{(2\pi)^6} g_n^0(\vec{k},\omega) \times$$
$$g_n^0(\vec{q}-\vec{k},-\omega) \exp[\,i\vec{q}(\vec{r'}-\vec{r}) + ie(x+x')(y'-y)H_0].$$
$$U_{n\vec{k}}(\vec{r'}) U_{n\vec{k}}^*(\vec{r})U_{n\vec{q}-\vec{k}}(\vec{r'}) U_{n\vec{q}-\vec{k}}^*(\vec{r}), \qquad (2.4)$$

where

$$g_n^0(\vec{k},\omega) = [i\omega - \xi_n(k)]^{-1}, \qquad (2.5)$$

ω is the Matsubara frequency, and $\xi_n$ is the energy of an electron in the *n*th band, $U_{n\vec{k}}^*(\vec{r})$ is the Bloch amplitude.

We shall perform averaging in equation (2.4) over the unit cell in order to exclude the rapid oscillations of the function $\Delta_n^*(\vec{r})$ (due to the Bloch functions) from consideration and to keep only the dependence on the coordinates of these functions due to the presence of an external magnetic field.

After that, integrating over the energy, we reduce equation (2.4) to the form

$$\Delta_m^*(\vec{r}) = \sum_n V_{mn} \frac{1}{\beta}\sum_\omega \int d\Omega \int \frac{d^3q}{(2\pi)^3} \frac{N_n \omega}{2|\omega|} \frac{i}{2i\omega + \vec{q}\vec{v}_n}$$
$$\int d^3r' \Delta_n(\vec{r'}) \exp[i\vec{q}(\vec{r}-\vec{r'}) + ieH_0(x+x')(y'-y)], \qquad (2.6)$$

where $N_n$ and $v_n$ are the density of electron states and the electron velocity on the *n*th cavity of the Fermi surface, respectively.

Thereupon, we calculate the right-hand side of equation (2.6) in accordance with the sophisticated technique of Maki and Tsuzuki [26] (see Appendix A). As a result, we obtain the set of equations

$$\Delta_m^*(x) = \sum_n \frac{V_{nm}\pi N_n}{\beta v_n} \int_1^\infty \frac{du}{u} \int_{-\infty}^\infty dx' \frac{\Delta_n^*(x') J_0[(x^2 - x'^2)eH_0\sqrt{u^2-1}]}{sh\frac{2|x-x'|\pi u}{\beta v_n}} \Theta(|x-x'| - \delta_n) \quad (2.7)$$

Here $J_0$ is the Bessel function.

The introduction of quantities $\delta_n$ corresponds to the cut-off of the electron-electron interaction in momentum space in the vicinity of the $n$th cavity of the Fermi surface. It is easy to see that at $T = T_c$ ($H_{c2} = 0$) equation (2.7) passes into the equation for determining the critical temperature of a superconducting transition.

Based on the last equation, we determine the quantity $\delta_n$

$$\delta_n = \frac{v_n}{2\gamma e_0 \omega_D^{(n)}}. \quad (2.8)$$

Here $e_0$ is the base of the natural logarithm, $\omega_D^{(n)}$ is the Debye frequency, and $\gamma$ is the Euler constant.
The solutions to equations (2.7) are as follows:

$$\Delta_m^*(x) = \Delta_m^* e^{-eH_0 x^2}. \quad (2.9)$$

Using (2.9), we reduce equation (2.7) to the form

$$\Delta_m^* = \sum_n V_{nm} N_n \rho_n^{-1/2} \Delta_n^* \int_1^\infty \frac{du}{u} \int_{\delta_n'}^\infty d\zeta \, exp^{-\frac{\zeta^2}{4}(u^2+1)} \frac{I_0[\frac{\zeta^2}{4}(u^2-1)]}{sh[u\zeta \rho_n^{-1/2}]}, \quad (2.10)$$

where

$$\delta_n' = (eH_0)^{1/2} \delta_n; \quad \rho_n = \frac{v_n^2 eH_0}{(2\pi T)^2}. \quad (2.11)$$

$I_0$ is the Bessel function of the imaginary argument, $e$ is the electron charge.
Equation (2.10) is easily reduced to the form

$$-\Delta_m^* + \sum_n V_{nm} N_n \Delta_n^* \ln \frac{2\gamma \omega_D^{(n)}}{\pi T_c} + \sum_n V_{nm} N_n \Delta_n^* \left[ \ln \frac{T_c}{T} - f(\rho_n) \right] = 0, \quad (2.12)$$

where

$$f(\rho_n) = \rho_n^{-1/2} \int_0^\infty d\zeta \int_1^\infty \frac{du}{u \, sh[u\zeta \rho_n^{-1/2}]} \left\{ 1 - exp\left[-\frac{\zeta^2}{4}(u^2+1)\right] I_0\left[\frac{\zeta^2}{4}(u^2-1)\right] \right\}. \quad (2.13)$$

The equality to zero of the determinant of system (2.12) corresponds to the presence of non zero solutions, that is, the formation of connected pairs. The field in the presence of which such solutions can appear is the upper critical field $H_{c2}$. So, the $H_{c2}$ value is determined from the condition of solvability of system (2.12):

$$af(\rho_1)f(\rho_2) + B_1 f(\rho_1) + B_2 f(\rho_2) + C = 0, \quad (2.14)$$

where

$$B_n = N_n V_{nn} - a \xi_C^{(n)}; \quad (n = 1,2)$$
$$C = 1 - N_1 V_{11} \xi_T^{(1)} - N_2 V_{22} \xi_T^{(2)} + a \xi_T^{(1)} \xi_T^{(2)}; \quad a = N_1 N_2 (V_{11} V_{22} - V_{12} V_{21}); \quad (2.15)$$
$$\xi_T^{(n)} = \ln \frac{2\gamma \omega_D^{(n)}}{\pi T}; \quad \xi_C^{(n)} = \ln \frac{2\gamma \omega_D^{(n)}}{\pi T_c}. \quad (2.16)$$

**Analytical Determination of the Upper Critical Field $H_{c2}$**

The analytical solutions to equation (2.14) could be computed for two limit cases as follows:
(a) $\rho_n \ll 1$ ($T_c - T \ll T_c$);        (b) $\rho_n \gg 1$ ($T \ll T_c$),

for which the functions f($\rho_n$) are defined in [26]:

$$f(\rho_n) = \frac{7}{6} \zeta(3) \rho_n - \frac{31}{10} \zeta(5) \rho_n^2 + \frac{381}{28} \zeta(7) \rho_n^3 + \cdots \quad at \quad \rho_n \ll 1 \quad (2.17)$$

$$f(\rho_n) = \ln \frac{2(2\gamma \rho_n)^{1/2}}{e_0} - \frac{1}{\pi^2 \rho_n} \left[ \zeta'(2) + \frac{\zeta(2)}{2} \ln \frac{2}{\pi^2 \gamma \rho_n} \right] 1 \ldots at \quad \rho_n \gg 1 \quad (2.18)$$

In the **(a)** case ($T_c - T \ll T_c$), applying the formulas in (2.14) and (2.17), we obtain the following expression for the $H_{c2}$ value:

$$\frac{H_{c2}(T)}{H_{c2}(0)} = \frac{8\gamma v_1 v_2}{e_0} \left[ v_1^2 \eta_1 + v_2^2 \eta_2 \right]^{-1} \frac{6}{7\zeta(3)} \Theta \times$$

$$\times \left[ 1 + \Theta \left\{ \frac{\lambda^2 \eta_1 + \frac{1}{\lambda^2}\eta_2}{\lambda \eta_1 + \frac{1}{\lambda}\eta_2} \frac{31}{10} \zeta(5) \left(\frac{6}{7\zeta(3)}\right) 2 - \frac{3}{2} \right\} \right] e^{\nu(\lambda) - \nu(1)}, \quad (2.19)$$

where

$$\lambda = v_1/v_2, \quad \theta = 1 - \frac{T}{T_c}, \quad \eta_1 = \frac{1}{2}(1+\eta), \quad \eta_2 = \frac{1}{2}(1-\eta),$$

$$\nu(\lambda) = \left[(\ln \lambda - \eta)^2 + \frac{4 N_1 N_2 V_{12} V_{21}}{a}\right]^{1/2}$$

$$\eta = \frac{N_1 V_{11} - N_2 V_{22}}{\sqrt{(N_1 V_{11} - N_2 V_{22})^2 + 4 N_1 N_2 V_{12} V_{21}}}. \quad (2.20)$$

In the **(b)** case, we obtain

$$H_{c2}(0) = \frac{\pi^2 T_c^2 e_0^2}{2 \gamma e v_1 v_2} \exp\left[\nu(1) - \nu(\lambda)\right]. \quad (2.21)$$

Given the condition $V_{22} = 0$, the expression for $H_{c2}$ in the low temperature range ($T \to 0$) may be reduced to the form

$$\frac{H_{c2}(T \to 0)}{H_{c2}(0)} = 1 + \frac{16}{\pi^2 e_0^2} \left(\frac{T}{T_c}\right)^2 \Phi(T) exp[\nu(\lambda) - \nu(1))], \quad (2.22)$$

where

$$\Phi(T) = \left(\lambda \gamma^- + \frac{1}{\lambda} \gamma^+\right) \left[\xi'(2) + \frac{1}{2} \xi(2) \left(2 \ln \frac{T}{T_c} + 2 \ln \frac{4}{\pi e_0} + \nu_0(\lambda) - \nu_0(1) + \ln \lambda\right)\right],$$

$$\nu_0(\lambda) = \nu(\lambda) \ at \ V_{22} = 0, \quad \gamma^{\pm} = \frac{1}{2}\left[1 \pm \left(\frac{\ln \lambda + \lambda_{11} \lambda_{12}^{-2}}{\nu(\lambda)}\right)\right]. \quad (2.23)$$

If we assume $v_1 = v_2$ in formulas (2.19-2.22), we obtain the respective relation for a single-band superconductor [25, 26].

Figure 1 depicts the dependence of the relation $H_{c2}(T)/H_{c2}(0)$ for a two-band isotropic superconductor on the basis of the above solutions to (2.19) and (2.22), using the following values of the constants of electron-electron interaction:

$\lambda_{11} = 0.3$; $\lambda_{12} = 0.12$; $N_1/N_2 = 0.8$.

Curves *1*, *2*, and *3* in this figure correspond to the cases $\frac{V_1}{V_2} = 1, 2,$ and 3, respectively. The dashed curve corresponds to the experimental dependence [37].

Using other values of the $\lambda_{nm}$ parameters of theory, we do not obtain a crucially new result for the relation $\frac{H_{c2}(T)}{H_{c2}(0)}$ for two-band isotropic superconductor at given values of the $v_1/v_2$ ratios.

A more important role is played by the values of ratios $v_1/v_2$. It is easy to see that with increasing $v_1/v_2$, the curvature in this dependence changes. The above results allow us to make a conclusion about a qualitative description of the experimental data by the behavior of the ratio $H_{c2}(T)/H_{c2}(0)$ as a function of temperature [37] in intermetallic compound $MgB_2$.

If we introduce the upper critical field $H_{c2}^0(0)$ and the critical temperature $T_{c0}$ of a low-temperature single-band superconductor, then, on the basis of (2.22), it is easy to obtain

$$\frac{H_{c2(0)}}{H_{c2}^0(0)} = \left(\frac{T_c}{T_{c0}}\right)^2 \frac{v_1}{v_2} \exp\{v(1) - v(\lambda)\}. \tag{2.25}$$

This formula implies that the upper critical field of a two-band superconductor $H_{c2}(0)$ can be several orders of magnitude higher than the value $H_{c2}^{(0)}$ of conventional single-band superconductors. Large values of $H_{c2}(0)$ are due to high $T_c$ and the relation $v_1/v_2 > 1$ or $v_1/v_2 \gg 1$. At the same time, the regard for the overlapping of energy bands in the determination of the value of $H_{c2}$ leads to a qualitatively new result in the dependence $H_{c2}(T)$: this dependence exhibits a positive curvature in the vicinity of the superconducting transition temperature (curves *2*, *3*) unlike a single-band superconductor (curve *1*).

For a more detailed description of the above theory, see [8, 9, 27, 33].

## 3. The Upper Critical Field $H_{c2}(ab)$ in a Two-Band Anisotropic Superconductor. Application to $M_gB_2$

Superconductivity in intermetallic compound $MgB_2$ at the temperature $T_c \sim 40$ K was discovered by a group of Japan researchers [23] in 2001. This discovery aroused a great interest because of relatively simple crystal structure of $MgB_2$ and high values of $T_c$, critical magnetic fields, and critical currents that allowed considering this material a high-temperature compound. Investigations showed that it was impossible to describe the high-temperature properties of $MgB_2$ by using the BCS-Bogoliubov-Gor'kov-Eliashberg theory of superconductivity, which was developed for isotropic systems. The $MgB_2$ compound is an anisotropic superconductor; the anisotropy is mainly shown in its band structure. According to the band structure of $MgB_2$ [38], several energy bands are overlapping on the Fermi surface of this material. Thus, two energy bands with different dimensions play the decisive role in superconductivity: the two-dimensional σ-band and the three-dimensional π-band (hereinafter, we shall term them band 1 and band 2, respectively).

In some researches [see 39, 40] it was recommended to apply the two-band model [2-5], which was proposed in 1959, to describe the superconducting properties of $MgB_2$. Our recent works [41, 42] confirm, in particular, that the above two-band model adequately describes the thermodynamic and magnetic properties of both pure compound $MgB_2$ and more complex systems in which Mg or B is replaced with other elements of the periodic table.

The experimental investigations of the magnetic properties of $MgB_2$ show the bright appearance of anisotropy of the upper critical field $H_{c2}$ [24]. The upper critical field $H_{c2}^{(ab)}$, which corresponds to an external magnetic field in the *(ab)* plane, is several times higher than the $H_{c2}^{(c)}$ value which corresponds to the magnetic field parallel to the *c*-axis

We set a problem to construct a microscopic theory of the upper critical field $H_{c2}$ of a pure anisotropic two-band superconductor, applicable in the entire temperature interval $0 < T < T_c$, to describe the pattern of the $H_{c2}$ value behavior as a function of temperature in $MgB_2$, to determine the curvature of the upper critical field $H_{c2}^{(ab)}$ and $H_{c2}^{(c)}$ in the vicinity of the superconducting transition temperature, and to reveal then the anisotropy of temperature dependence of the coefficient $\gamma_H = H_{c2}^{(ab)}/H_{c2}^{(c)}$. Also we shall try to determine the effect of the mechanism of occupation of energy bands on the values $T_c$ and $H_{c2}^{(ab)}$ upon doping of a system with electrons or holes.

The study of two-band systems is based on the microscopic approximation of the theory of superconductivity.

## The set of equations for the order parameters $\Delta_m^*$ in the magnetic field $\vec{H}$ directed parallel to the *(ab)* plane

We consider high values of the magnetic field $H$ (close to the value of the upper critical field $H_{c2}$), i.e., the subcritical region in the vicinity of the unstable normal state.

In this region, at $\vec{H} \to \vec{H}_0 = \vec{H}_{c2}$ the order parameters $\Delta_m \to 0$, and it is possible be restricted to the linear values of $\Delta_n^*$ terms in equation (2.1). For all this, we obtain

$$\Delta_m^*(\vec{r}) = \frac{1}{\beta} \sum_n V_{mn} \sum_\omega \int d\vec{r}' g_n(\vec{r}',\vec{r},\omega) \Delta_n^*(r') g_n(\vec{r}',\vec{r},-\omega) . \qquad (3.1)$$

In the presence of a magnetic field, the electronic Green function $g_n$ is defined by the expression [25]:

$$g_n(\vec{r},\vec{r}',\omega) = e^{i\varphi(\vec{r},\vec{r}')} g_n^0(\vec{r},\vec{r}',\omega). \qquad (3.2)$$

Here $g_n^0$ is the Green function in the normal state without magnetic field. The presence of a magnetic field is taken into consideration by the phase multiplier:

$$\varphi(\vec{r}',\vec{r}) = \int_{\vec{r}}^{\vec{r}'} A(\vec{l}) d\vec{l}. \qquad (3.3)$$

The set of equations (3.1) was obtained on the basis of the results of work [31] (see also (2.1)) in the diagonal approximation by the band's indices. This approximation takes into account the processes of superconducting pairing of electrons into each energy band and their tunneling as a whole from one band into another.

We shall substitute definition (3.2) in this formula, performing expansion of the function $g_n^0$ in terms of the Bloch functions; after that, we shall average this equation over the amplitudes of the Bloch functions. In this way, we have

$$\Delta_m^*(\vec{r}) = \sum_n V_{mn} \frac{1}{\beta} \sum_\omega \int d\vec{r}' \sum_{\vec{k}} \sum_{\vec{q}} g_n^0(\vec{k},\omega) g_n^0(\vec{q}-\vec{k},-\omega) \times$$
$$\times \Delta_n^*(\vec{r}') e^{2i\varphi(\vec{r}',\vec{r})} e^{i\vec{q}(\vec{r}'-\vec{r})} , \qquad (3.5)$$

$$g_n^0(\vec{k},\omega) = [i\omega - \xi_n(\vec{k})]^{-1}, \qquad (3.6)$$

where $\omega$ is the Matsubar frequency, $\xi_n$ is the electron energy in the *n*th band.

Consider the magnetic field $\vec{H}$ (which is parallel to the (*ab*) plane) directed along the *y*-axis. Herein, it is possible to choose $A_z = -\frac{H_0}{2}(x+x')$, $A_y = A_z = 0$ in the symmetric view, and on the basis of (3.3) we obtain

$$2\varphi(\vec{r}',\vec{r}) = eH_0(x+x')(z-z'). \qquad (3.7)$$

We shall represent the dispersion law for the σ- and π-bands (it is singed by index 1 and 2, respectively) in the form:

$$\xi_1(\vec{k}) = \zeta_1 + \frac{k_x^2 + k_y^2}{2m_1} + \frac{k_z^2}{2M} - \mu , \quad \xi_2(\vec{k}) = \zeta_2 + \frac{k_x^2 + k_y^2 + k_z^2}{2m_2} - \mu , \qquad (3.8)$$

where $M \gg m_1$. The satisfiability of this inequality means a weak deviation of the dispersion law of the first band from two-dimensionality. The Fermi surface of the other band is assumed spherical for simplicity.

We shall introduce the definition of function $g_n^0$ (3.6) in (3.5) and we shall pass from summation over $\vec{k}$ to integration over energy in each energy band in accordance with dispersion law (3.8) (over the cylindrical space in band 1 and the spherical space in band 2).

We shall consider the order parameters $\Delta_m^*(r)$ to depend on $x$ [25, 26]. Taking into account this dependence, we shall perform a series of operations in the set of equations (3.4): integration over energy $\xi_1^0(\vec{k}) = \xi_1(k)$ at $k_z = 0$ and $\xi_2$, partial integration over coordinate variables and over $\vec{q}$, and summation over $\omega$. The technique of calculations of this sort is given in Appendix A. In this way, we obtain (see [31])

$$\Delta_1^*(x) = \lambda_{11} \frac{2T}{v_1} \int_1^\infty \frac{du}{\sqrt{u^2-1}} \int_{-\infty}^\infty \frac{\Delta_1^*(x')dx'}{sh\frac{2\pi T|x-x'|u}{v_1}} \frac{\sin[\tilde\varepsilon e H_0 u(x^2-x'^2)]}{\tilde\varepsilon e H_0 u(x^2-x'^2)} \theta(|x-x'|-\delta_{11}) +$$

$$+\lambda_{12} \frac{\pi T}{v_2} \int_1^\infty \frac{du}{u} \int_{-\infty}^\infty \frac{dx'\Delta_2^*(x')}{sh\frac{2\pi T|\vec{x}-\vec{x}'|u}{v_2}} J_0\left[(x^2-x'^2)e H_0 \sqrt{u^2-1}\right]\theta(|x-x'|-\delta_{12}), \qquad (3.9)$$

$$\Delta_2^*(x) = \lambda_{21} \frac{T}{v_1} \int_1^\infty \frac{du}{\sqrt{u^2-1}} \int_{-\infty}^\infty \frac{\Delta_1^*(x')dx'}{sh\frac{2\pi T|x-x'|u}{v_1}} \frac{\sin[\tilde\varepsilon e H_0 u(x^2-x'^2)]}{\tilde\varepsilon H_0(x'-x'^2)} \theta(|x-x'|-\delta_{11}) +$$

$$+ \lambda_{22} \frac{\pi T}{v_2} \int_1^\infty \frac{du}{u} \int_{-\infty}^\infty dx'\Delta_2^*(x') \frac{J_0[(x^2-x'^2)eH_0\sqrt{u^2-1}]}{sh\frac{2\pi T|x-x'|u}{v_2}} \theta(|x-x'|-\delta_{22}), \qquad (3.10)$$

where $\tilde\varepsilon = \frac{p_0 \varepsilon}{m_1 v_1}$, $p_0$ is an impulse of the cylinder cutoff along the $z$ axis, $\lambda_{nm}$ is the renormalized constants of the two-band theory related with the strong electron-phonon coupling and Coulomb electron-electron interaction [41, 42].

Every term of the right-hand part of the set of equations (3.9) and (3.10) contains the step function $\theta(|x-x'|)-\delta_{ij})$, which corresponds to the trimming of the integrals in the impulse space by definition of the parameters $\delta_{ij}$ [26, 36].

If the interband interaction is neglected, what can be done when $\lambda_{11}, \lambda_{22} \gg \lambda_{12}, \lambda_{21}$, then on the basis of (3.9) and (3.10) we shall obtain two independent equations, from which it follows that, at $\tilde\varepsilon \to 0$, $\Delta_1^*(x)$ does not depend on the magnetic field value. This fact will be used for choosing the solutions for the order parameters $\Delta_1^*(x)$ and $\Delta_2^*(x)$ conformably to MgB$_2$, inasmuch as the interband electron-phonon interactions are low for this compound. This circumstance allows choosing the solution for parameters $\Delta_1^*(x)$ and $\Delta_2^*(x)$ in the form:

$$\Delta_1^*(x) = \Delta_1^* e^{-\tilde\varepsilon e H_0 x^2}, \quad \Delta_2^*(x) = e^{-e H_0 x^2} \qquad (3.11)$$

In this way, $\lim_{\varepsilon \to 0} \Delta_1^*(x) = \Delta_1^*$.

We shall put the values of (3.11) into the set of equations (3.9) and (3.10), then we shall multiply equation (3.9) by $e^{-\tilde\varepsilon e H_0 x^2}$ and (3.10) by $e^{-e H_0 x^2}$; after that, we shall integrate over variable $x$ in the infinite limits in the two equations. Then, passing to dimensionless variables like $\xi = (\tilde\varepsilon e H_0)^{1/2} x$ and $\xi' = (\tilde\varepsilon e H_0)^{1/2} x'$ and performing a series of other transformations (see [31], appendix A), we shall reduce the set of equations (3.9) and (3.10) to the form:

$$\Delta_1^* = \Delta_1^* \lambda_{11} \xi^{(1)}(T_c) + \Delta_1^* \lambda_{11} [\xi^{(1)}(T) - \xi^{(1)}(T_c)] - \lambda_{11} f_{11}(\rho_1 \tilde\varepsilon) \Delta_1^* +$$

$$+\tilde{\lambda}_{12}\Delta_2^*\xi^{(2)}(T_c)+\tilde{\lambda}_{12}[\xi^{(2)}(T)-\xi^{(2)}(T_c)]\Delta_2^* - \tilde{\lambda}_{12} f_{12}(p_2\tilde{\varepsilon})\Delta_2^*, \tag{3.12}$$

$$\Delta_2^* = \tilde{\lambda}_{21}\Delta_1^*\xi^{(1)}(T_c) + \tilde{\lambda}_{21}\Delta_1^*[\xi^{(1)}(T)-\xi^{(1)}(T_c)] - \tilde{\lambda}_{21} f_{21}(\rho_1\tilde{\varepsilon})\Delta_1^* +$$
$$+\lambda_{22}\Delta_2^*\xi^{(2)}(T_c)+\lambda_{22}[\xi^{(2)}(T)-\xi^{(2)}(T_c)]-\lambda_{22} f_{22}\,\rho_2\,\tilde{\varepsilon})\Delta_2^*. \tag{3.13}$$

where

$$\xi^{(n)}(T)=\int_{-d_n}^{d_{cn}} d\varepsilon\,\frac{\text{th}\frac{\beta\varepsilon}{2}}{2\varepsilon},\ \xi^{(n)}(T_c)=\int_{-d_n}^{d_{cn}} d\varepsilon\,\frac{\text{th}\frac{\beta_c\varepsilon}{2}}{2\varepsilon},\ \tilde{\lambda}_{12}=\tilde{\varepsilon}^{1/2}\sqrt{\frac{2}{\tilde{\varepsilon}+1}}\lambda_{12},\ \tilde{\lambda}_{21}=\sqrt{\frac{2}{\tilde{\varepsilon}+1}}\lambda_{21}, \tag{3.14}$$

$\beta = 1/T$, and values $d_n = \mu - \zeta_n$, $d_{cn} = \zeta_{cn} - \mu$ are the parameters of the integrals taken over energy at variable density of charge carriers. Under phonon mechanism of superconductivity (the MgB$_2$ case) these parameters have the form:

$$d_n = \begin{cases}\mu-\zeta_n & \text{at}\quad \mu-\zeta_n \le \omega_D^{(n)} \\ \omega_D^{(n)} & \text{at}\quad \mu-\zeta_n > \omega_D^{(n)}\end{cases} \qquad d_{cn}=\begin{cases}\omega_D^{(n)} & \text{at}\quad \zeta_{cn}-\mu > \omega_D^{(n)} \\ \zeta_{cn}-\mu & \text{at}\quad \zeta_{cn}-\mu < \omega_D^{(n)}\end{cases} \tag{3.15}$$

Here $\omega_D^{(n)}$ is the characteristic phonon frequency corresponding to the $n$th energy band. The functions $f_{nm}$ containing the dependence on the magnetic field are as follows:

$$f_{11}=\frac{(\tilde{\varepsilon}\rho_1)^{-1/2}}{\pi}\int_{-1}^{1} dy \int_1^\infty \frac{du}{\sqrt{u^2-1}}\int_0^\infty \frac{d\zeta}{\text{sh}\frac{\zeta u}{(\tilde{\varepsilon}\rho_1)^{\frac{1}{2}}}}\left(1-\exp\left[-\frac{\zeta^2}{2}(1+u^2 y^2)\right]\right) \tag{3.16}$$

$$f_{12} = \rho_2^{-1/2}\int_0^\pi d\varphi \int_1^\infty \frac{du}{u}\int_0^\infty \frac{d\zeta}{\text{sh}(\zeta u/\rho_2^{1/2})}\left(1-e^{\frac{-\zeta^2(1+\tilde{\varepsilon})}{4}\left[1-\left(\frac{\tilde{\varepsilon}-1+2i\sqrt{u^2-1}\cos\varphi}{1+\tilde{\varepsilon}}\right)^2\right]}\right) \tag{3.17}$$

$$f_{21}=\frac{\rho_1^{-1/2}}{\pi}\int_{-1}^1 dy \int_1^\infty \frac{du}{\sqrt{u^2-1}}\int_0^\infty \frac{d\zeta}{\text{sh}\frac{\zeta u}{\rho_1^{1/2}}}\left(1-e^{\frac{-\zeta^2(1+\tilde{\varepsilon})}{4}\left[1-\left(\frac{1-\tilde{\varepsilon}-2i\tilde{\varepsilon}uy}{1+\tilde{\varepsilon}}\right)^2\right]}\right)$$

(3.18)

$$f_{22}=\rho_2^{-1/2}\int_1^\infty \frac{du}{u}\int_0^\infty \frac{d\zeta}{\text{sh}\frac{\zeta u}{\rho_2^{1/2}}}\left[1-e^{-\frac{\zeta^2}{4}(u^2+1)} I_0\left(\frac{\zeta^2(u^2-1)}{4}\right)\right]. \tag{3.19}$$

The dimensionless parameter $\rho_n^{1/2}=\dfrac{v_n(eH_0)^{1/2}}{2\pi T}$, which contains the value $H_0 = H_{c2}^{(ab)}$, was introduced in the definition of functions $f_{nm}$, $v_n$ is the velocity of electrons on the $n$th cavity of the Fermi surface, $I_0$ is the Bessel function of the imaginary argument.

**Determination of the upper critical field $H_{c2}(ab)$. The analytical solutions**

From the condition of solvability the set of equations (3.12) and (3.13) at $H \to \infty$, we obtain the equation for the superconducting transition temperature $T_c$:

$$a\,\xi^{(1)}(T_c)\xi^{(2)}(T_c)-\lambda_{11}\xi^{(1)}(T_c)-\lambda_{22}\xi^{(2)}(T_c)+1=0, \tag{3.20}$$

where $a = \lambda_{11} \lambda_{22} - \lambda_{12} \lambda_{21}$, and $\xi^{(n)}(T)$ is given by formula (3.14). Equating the determinant of set (3.12), (3.13) to zero and using equation (3.20), we obtain the equation for determination of the upper critical field $H_0 = H_{c2}^{(ab)}$ with which the superconducting pairs appear. This equation has the form:

$$\lambda_{11} \lambda_{22} \tilde{F}_{11} \tilde{F}_{22} - \lambda_{12} \lambda_{21} \tilde{F}_{12} \tilde{F}_{21} + \lambda_{11}[1 - \lambda_{22} \xi^{(2)}(T_c)] \tilde{F}_{11} + \lambda_{22}[1 - \lambda_{11} \xi^{(1)}(T_c)] \times$$
$$\times \tilde{F}_{22} + \lambda_{12} \lambda_{21} \xi^{(2)}(T_c) \tilde{F}_{21} + \lambda_{21} \lambda_{12} \xi^{(1)}(T_c) \tilde{F}_{12} = 0, \tag{3.21}$$

where

$$\tilde{F}_{mn} = f_{mn} + \xi^{(n)}(T_c) - \xi^{(n)}(T), \tag{3.22}$$

$$f_{mn} = f_{mn}(\rho_n, \tilde{\varepsilon}), \qquad \xi^{(n)}(T_c) - \xi^{(n)}(T) = -\ln T_c / T. \tag{3.23}$$

For investigating the superconducting properties of the system with a variable density of charge carriers, it is necessary to supplement equation (3.21) with the correlation that defines the chemical potential μ (the charge carrier density $\tilde{n}$) [7, 18, 35]:

$$\tilde{n} = \sum_m N_m \int_{-d_m}^{d_{cm}} d\varepsilon_m \left[ \frac{E_m(\vec{k}) - |\varepsilon_m(\vec{k}) - \mu|}{E_m(\vec{k})} + \frac{2|\varepsilon_m(\vec{k}) - \mu|}{E_m(\vec{k})} \frac{1}{1 + \exp \beta E_m(\vec{k})} \right] \tag{3.24}$$

Considering $d_m/T_c$ and $d_{mc}/T_c \gg 1$, on the basis of correlation (3.24), we obtain

$$\tilde{n} = \sum_n Nn[\zeta_{cn} - \zeta_n - |\zeta_{cn} - \mu| + |\zeta_n - \mu|]. \tag{3.25}$$

The set of equations (3.20), (3.21), and (3.25) allows determining the value of the critical temperature $T_c$ and the upper critical field $H_{c2}^{(ab)}(T)$ at any density of charge carriers. If we set the chemical potential (the charge carrier density $\tilde{n}$) at the level of its value for pure MgB$_2$ (e.g., $\mu = \mu_0$, $\tilde{n} = \tilde{n}_0$), then the aforesaid set of equations makes it possible to determine the temperature of the superconducting transition $T_c$ and the upper critical field $H_{c2}^{(ab)}$ of pure MgB$_2$.

In this work, we shall analyze the case of pure MgB$_2$.

Equation (3.21) contains complex integral dependences $f_{mn}$ (3.16) – (3.19), it is possible to solve this equation in the entire temperature interval $0 < T < T_c$ by only numerical methods. However, it is possible to find the analytical solutions to this equation in two limit cases: in the vicinity of the critical temperature $\rho_n \ll 1$, when $T_c - T \ll T_c$ and in the range of low temperatures $\rho_n \gg 1$, at $T \ll T_c$.

To this effect, for each of these cases, we obtain asymptotic expressions for functions $f_{nm}$, having performed an expansion in the parameter $\rho_n \ll 1$ (case a) and in $\rho_n^{-1} \ll 1$ (case b). We shall substitute values of these functions into (3.21) with some preliminary simplifications (see Appendix B). Thereupon, we obtain

a) $(T_c - T \ll T_c)$, $\rho_n \ll 1$:

$$\rho_c^0 = \frac{v_1 v_2 e H_0(T)}{(2\pi T_c)^2} = \left[ \frac{v\theta}{\tilde{\beta}} + \frac{v(\tilde{\beta}^2 - 2\tilde{\alpha} v) \theta^2}{\tilde{\beta}^2} \right] \lambda^{-1} \frac{T}{T_c} \tag{3.26}$$

where

$$\theta = 1 - \frac{T}{T_c}, \qquad \lambda = v_1 / v_2 \tag{3.27}$$

b) $T_c \ll T$, $\rho_n \gg 1$

$$\rho_c^0 = \frac{e_0}{4\gamma} e^{-\frac{1}{2}\eta_- - \frac{3}{2}v(1) + \Omega(\lambda)} \left[ 1 - \frac{F(T)}{\Omega(\lambda)} \right] \tag{3.28}$$

The determination of other values in (3.26) and (3.27) see in Appendix B.

## 4. Case $\vec{H} \parallel c$. The determination of $H_{c2}^{(c)}$

Above, we have described the theory of the upper critical field H$_{c2}$(ab) for a two-band anisotropic system. In MgB$_2$, this value corresponds to the maximal value of the upper critical field. It is of interest to adduce the results of calculations of $H_{c2}^{(c)}$, which corresponds to the minimal value of this quantity in the intermetallic $MgB_2$ when the magnetic field $\vec{H} \parallel \vec{c}$. This information, together with the data on the behavior adduced for $H_{c2} \parallel (ab)$, will allow obtaining the temperature dependence of the coefficient $\gamma_H = H_{c2}^{(ab)}/H_{c2}^{(c)}$, which determines the anisotropy for the upper critical field.

We consider $\vec{H} \parallel \vec{c}$. In this case, it is possible to choose $A_x = A_z = 0$, $A_y = H_0(x+x')/2$ and we have for the phase multiplier:
$$2\varphi(\vec{r}',\vec{r}) = eH_0(x+x')(y'-y).\tag{4.1}$$

In the case of **H** || **c** considered here, the average velocity of electrons in the *ab* plane plays an important role for either energy band, while the value of velocity in the *z* direction is immaterial. This makes it unnecessary to introduce a parameter controlling the deviation of the σ-band from the 2D behavior and renders the problem less anisotropic (for details, see [34]). Let us consider some results. The critical magnetic field $H_{c2}^{(c)}$ is defined by the equation
$$a\tilde{f}_1(\rho_1)\tilde{f}_2(\rho_2) + \overline{\lambda_{11} - a\xi_c}\ \tilde{f}_1(\rho_1) + \overline{\lambda_{22} - a\xi_c}\ \tilde{f}_2(\rho_2) = 0,\tag{4.2}$$

where

$$\tilde{f}_1(\rho_1) = f_1(\rho_1) - \ln\left(\frac{T_c}{T}\right)^{1/2}, \tilde{f}_2(\rho_2) = f_2(\rho_2) - \ln\frac{T_c}{T}\tag{4.3}$$

$$f_1(\rho_1) = \frac{\rho_1^{-1/2}}{\pi}\int_1^\infty \frac{du}{\sqrt{u^2-1}}\int_0^\infty \frac{d\xi}{\sinh\frac{\xi u}{\rho_1^{1/2}}}\left[1 - e^{\frac{-\xi^2 u^2}{2}}\right].\tag{4.4}$$

The function $f_2(\rho_2)$ corresponds to the expression for $f_{22}$ (3.19).

The asymptotes for the function $f_1(\rho_1)$ are given by

$$f_1(\rho_1) = \frac{7}{8}\varsigma(3)\rho_1 - \frac{31\cdot 3}{32}\varsigma(5)\rho_1^2 + \cdots \text{ at } \rho_1 \ll 1,$$

$$f_1(\rho_1) = \frac{1}{4}\ln 2\gamma\rho_1 + \frac{1}{\rho_1}\frac{3}{2\pi^2}\varsigma(2) - \frac{1}{\rho_1^2}\frac{9}{4\pi^4}\varsigma(4) + \cdots \text{ at } \rho_1 \gg 1.\tag{4.5}$$

The asymptotes for the function $f_2(\rho_2)$ coincide with formulas (B.1) and (B.2).

The solution to equation (4.2) with regard for the aforesaid asymptotes for the functions $f_n(\rho_n)$ allows obtaining the solution in the vicinity of the critical temperature $(T_c - T) \ll T_c$, $\rho_n \ll 1$ and in the vicinity of the zero temperature $(T \ll T_c)$, $\rho_n \gg 1$. These solutions are given by, respectively,

At $\rho_n \ll 1$,
$$\rho_1(T) = \frac{v_1^2\, eH_0(T)}{(2\pi T_c)^2} = \alpha_1\theta + \alpha_2\theta^2,\tag{4.6}$$

$$\alpha_1 = \frac{\eta_1 + 2\eta_2}{7\zeta(3)\left[\frac{\eta_1}{4} + \frac{\eta_2}{3}\frac{1}{\lambda^2}\right]},$$

$$\alpha_2 = \alpha_1 \left\{ \frac{\left[\frac{31\cdot 3}{16}\eta_1 + \frac{31}{5}\frac{1}{\lambda^2}\eta_2\right]\zeta(5)\alpha_1}{\left[\frac{1}{4}\eta_1 + \frac{1}{3}\frac{1}{\lambda^2}\eta_2\right]7\zeta(3)} + 1 \right\}., \quad (4.7)$$

here

$$\theta = 1 - T/T_c, \quad \eta_{1,2} = \frac{1}{2}[1 \pm \eta], \quad \eta = [(\lambda_{11} - \lambda_{22})^2 + 4\lambda_{11}\lambda_{22}]^{-1/2}(\lambda_{11} - \lambda_{22}). \quad (4.8)$$

At $\rho_n \gg 1$,

$$\rho_c^0(T) = \frac{v_1 v_2\, eH_0(T)}{(2\pi T_c)^2} = \frac{e_0}{4\gamma} e^{\frac{1}{2}\eta_- - \frac{3}{2}\nu(1) + \Omega(\lambda)}\left[1 - \frac{F_\parallel^C(T)}{\Omega(\lambda)}\right] \quad (4.9)$$

$$\eta_- = \frac{\lambda_{11} - \lambda_{12}}{a}, \quad \nu(\lambda) = \frac{1}{a}\sqrt{(\lambda_{11} - \lambda_{22} - a\ln\lambda)^2 + 4\lambda_{12}\lambda_{21}}, \quad (4.10)$$

$$\Omega(\lambda) = \left\{\left[-\ln\frac{e_0\lambda}{2} + \frac{3}{2}\eta - \frac{1}{2}\nu(1)\right]^2 + \frac{8\lambda_{12}\lambda_{21}}{a^2}\right\}. \quad (4.11)$$

We determine $H_{c2}^{(c)}$ by expression (4.6) in the vicinity of the critical temperature $T_c - T \ll T_c$, and formula (4.9) defines the same quantity in the vicinity of the zero temperature.

### 5. Application of the theory to MgB$_2$. Numeric computation and discussion of the results. Anisotropy of the upper critical field

The behavior of the upper critical field $H_{c2}^{(ab)}$ and $H_{c2}^{(c)}$ as a function of correlation $\frac{T}{T_c}$ was investigated in this work on the fundamental principles of the theory of superconductivity in a magnetic field [25, 26, 31, 32] in the entire temperature interval $0 < T < T_c^0$.

An anisotropic system equivalent to MgB$_2$ was considered in the magnetic field parallel to the (ab) plane: there were two overlapping energy bands with different dimensions on the Fermi surface (almost two-dimensional σ-band with a weak dispersion in the direction of the z axis and the three-dimensional π-band for which the Fermi surface was chosen in the form of a sphere).

On the basis of the Ginzburg-Landau equations, generalizing the calculation method of Maki and Tsuzuki for the two-band case and taking into account the above features of the band structure inherent in MgB$_2$, we obtain the set of equations for determining $T_c$ (3.19) and the upper critical field $H_{c2} = H_{c2}^{(ab)}$ (3.21) in the magnetic field parallel to the (ab) plane.

These equations can be used to study either pure MgB$_2$ or more complex systems obtained by partial replacement of Mg и B with other chemical elements.

Note that the considered set of equations (3.20) and (3.21) does not contain dissipation of electrons on the impurity potential; however, it takes into account the mechanism of filling of energy bands upon introduction of electrons or holes.

The asymptotes of the functions $f_{mn}$ (B.1), (B.2), which are engaged into the definition of equation (3.21) were found and the analytical solutions for $H_{c2}^{(ab)}$ were obtained in the vicinity of the critical temperature $T_c - T \ll T_c$, ($\tilde{\varepsilon}\rho_n \ll 1$) (3.26) and in the vicinity of the zero temperature $T \ll T_c$, ($\tilde{\varepsilon}\rho_n \gg 1$) (3.10). The analytical solutions for the critical field $H_{c2}^{(c)}$, which corresponds to the direction of the magnetic field parallel to the *z* axis, were obtained earlier [34], and brief results were adduced in section 4 of this work.

There is a great spread in values of the electron-phonon interaction parameters in the literature. Therefore, we adduced the numeric results using two sets of these parameters:
a) $\lambda_{11} = 0{,}302$; $\lambda_{22} = 0{,}135$; $\lambda_{12} = 0{,}04$; $\lambda_{21} = 0{,}038$ [41, 42].
b) $\lambda_{11} = 0{,}362$; $\lambda_{22} = 0{,}255$; $\lambda_{12} = 0{,}174$; $\lambda_{21} = 0{,}165$.

The important role in the considerable difference of $H_{c2}(ab)$ from $H_{c2}(c)$ is played by the strong anisotropy of the electron energy spectrum, which exhibits a weak dispersion of the electron energy along the *z* axis (low values of the average velocity of electrons in this direction). In our theory, this circumstance is defined by the presence of a small parameter $\tilde{\varepsilon}$. The small value of the interband electron-electronic interaction also plays a considerable role.

The results of the numerical solutions are presented below in Figs. 2 and 3. The figure captions contain the values of the theory parameters used. For other parameters of the theory, see [35]. Figure 2 shows the temperature dependence of the critical fields $\rho_c^0(ab)$ and $\rho_c^0(c)$; Fig. 3, the temperature dependence of the anisotropy coefficient $\gamma_H = \frac{\rho_c^0(ab)}{\rho_c^0(c)}$. The selected parameters of the theory are listed in the figure captions. It follows from Fig. 2 that the anisotropy of the energy spectrum has a strong effect on the values of critical magnetic fields: $H_{c2}(ab) \gg H_{c2}(c)$. Figure 3 shows the significant temperature dependence of the anisotropy coefficient $\gamma_H = \frac{\rho_c(ab)}{\rho_c(c)}$.

It is possible to make a conclusion that the theory proposed in this paper, which describes the upper critical field as a function of temperature and also the anisotropy parameter in a two-band anisotropic superconductor ($MgB_2$), gives a qualitative conformity with experimental data [43]-[45].

## Two-band system with a variable density of charge carriers

As noted above, section 3 of this paper contains the basic equations that allow determining the behavior of the quantities $T_c$, $H_{c2}(ab)$, $H_{c2}(c)$, and $\gamma_H$ as a function of temperature and in a system with a variable density of charge carriers upon doping of the system with electrons or holes. In addition, we take into account the mechanism of filling of the energy bands upon replacement of atoms of Mg or B with other elements of the periodic table.

Let us consider the results of numerical calculations obtained for the upper critical field $H_{c2}^{(ab)}$ using Eq.(3.21) and relation (3.25) defining the chemical potential as well as Eq.(4.21) for upper critical field $H_{c2}^{(c)}$ (for detail, see [34]).

We shall use the electron–phonon interaction constant corresponding to $MgB_2$ (see Sect. 5) as well as $\lambda = v_1/v_2 = 0.8$. We also assume that the value of chemical potential in undoped $MgB_2$ is $\mu_0 = 0.74$ eV [47]. In all figures considered here, parameter $\tilde{\varepsilon} = 0.31$. This value yields the results closest to experimental values.

Figure 4 shows the temperature dependences of the upper critical fields $H_{c2}^{(ab)}$ and $H_{c2}^{(c)}$ for pure $MgB_2$ (curves 1 and 1′, respectively) and quantities $\widetilde{H}_{c2}^{(ab)}$ and $\widetilde{H}_{c2}^{(c)}$ for electron-doped system $MgB_2$ (curves 2 and 2′, respectively). We can see that $\widetilde{H}_{c2}^{(ab)} \gg \widetilde{H}_{c2}^{(c)}$. This result is in good agreement with the results of many theoretical publications as well as with experimental data. Strong anisotropy of the upper critical field is due to the weak electron energy dispersion at the Fermi surface in the direction of the *c* axis and the small value of the average velocity of electrons on the Fermi surface in this direction. In the case when other elements facilitating doping with electrons (an increase in chemical potential) are substituted for Mg and B atoms, the behavior of the upper critical field as a function of temperature (curves 2 and 2$'$) is analogous to the case of pure $MgB_2$. However, the values of these quantities are lower than for pure $MgB_2$. A correlation is observed between superconducting transition temperature $T_c$ and the values of the upper critical field upon an increase in the chemical potential. An increase in the hole conductivity, however, does not affect the values of the superconducting transition temperature and the upper critical field.

Figure 5 shows the dependence of the ratio of the superconducting transition temperatures of doped and pure $MgB_2$ (curve *1*) and the critical fields $H_{c2}$(ab) (curve *2*) and $H_{c2}$(c) (curve *3*) at the temperature T = 0 on the electron density (chemical potential). We can see that for $\mu > 0.74$ eV all quantities decrease with increasing electron density of charge carriers, remaining constant for $\mu < 0.74$ eV. Consequently, hole doping does not affect the superconducting transition temperature and the upper critical field.

Figure 6 shows the temperature dependence of anisotropy coefficient in pure $MgB_2$ ($\mu_0 = 0.74$ eV) and in doped $MgB_2$ ($\mu = 0.76$ eV). The circles correspond to the experimental data borrowed from [45].

These results are in satisfactory agreement with experimental data on the magnetic properties of intermetallic compound $MgB_2$ (both pure and doped with electrons and holes), which indicates the effectiveness of our two-band model in describing the properties of real materials and in calculating the anomalies in physical properties associated with anisotropy of the system.

It should be noted that the main mechanism of action of a substitutional impurity on the system was considered here as the effect of energy band filling, on the assumption that scattering from the impurity potential is weak. The allowance for electron scattering considerably complicates the results for systems whose impurity scattering is strong.

## Conclusions

Section 1 of this work, as well as some references, indicates a greater number of anomalies in the behavior of the superconducting properties of intermetallic compound $MgB_2$. These anomalies cannot be explained on the basis of the BCS theory intended for describing the properties of isotropic superconductors. To understand the appearance of these anomalies, in this paper, we construct a microscopic theory of the upper critical field $H_{c2}$ in a system with the anisotropic energy spectrum. The case in hand is the regard for the overlapping of energy bands on the Fermi surface. These bands may be equivalent with the same topology of the Fermi surface cavities or with different topologies (e.g., when the energy bands have different dimensions).

In this paper, we study the effect of anisotropy of the energy spectrum on the temperature dependence of the upper critical fields $H_{c2}$(ab) and $H_{c2}$(c) and the anisotropy coefficient $\gamma_H$ and the effect of replacement of atoms of Mg and B with other elements of the periodic table on these physical quantities and their temperature dependence. The features of the magnetic properties of $MgB_2$ are as follows.

(i) The appearance of the positive curvature in the temperature dependence of the upper critical field $H_{c2}$ in the vicinity of the superconducting transition temperature in two-band systems (in the one-band case, the curvature is negative). The positivity of curvature is due to different values of the average velocities of electrons in different cavities of the Fermi surface ($v_1 \neq v_2$).

(ii) There is the possibility of a significant increase in the upper critical field $H_{c2}$ in a two-band system in comparison with a one-band system due to high values of $T_c$ and to the inequality of the values of $v_1$ and $v_2$.

(iii) The value of $H_{c2}$ in pure $MgB_2$ exhibits a high degree of anisotropy; that is, $H_{c2}$(ab) is several times higher than $H_{c2}$(c) in a wide temperature range. This situation is explained by the weak dispersion of electron energy in the direction of the *z* axis and by the low values of the electron velocity in this direction because the σ-band is two-dimensional. Upon doping of MgB$_2$ with electrons, the values of H$_{c2}$ decrease as compared to the case of pure MgB$_2$.

(iv) The doping of compound MgB$_2$ with electrons, which contributes to an increase in the chemical potential, decreases the upper critical field H$_{c2}$ and the value of T$_c$ in comparison with the case of pure MgB$_2$. The doping with holes does not affect the superconducting transition temperature and the upper critical field H$_{c2}$. The doping of MgB$_2$ significantly changes the temperature dependence of the anisotropy coefficient $\gamma_H = \dfrac{H_{c2}(ab)}{H_{c2}(c)}$.

**Appendix A**

We shall represent (2.6) in the form

$$\Delta_m = \sum_n V_{mn} K_n,  \tag{A.1}$$

where K$_n$ is given by

$$K_n = \frac{N_n}{4}\frac{1}{\beta}\sum_\omega \int_{-1}^{1} d\cos\theta \int_0^{2\pi} d\varphi \int \frac{d\vec{q}}{(2\pi)^3}\frac{\omega}{2|\omega|(2i\omega+\vec{q}\vec{v}_n)} \int d\vec{r}'\, \Delta_n(\vec{r}') \times$$
$$\times \exp\{i\vec{q}(\vec{r}-\vec{r}') + ieH_0(x+x')(y'-y)\}.$$

1. We choose the polar axis in the *x* direction, i.e., we write
$$\vec{v}_n\vec{q} = v_n\,(q_x\cos\theta + q_y\sin\theta\cos\varphi + q_z\sin\theta\sin\varphi)\;.$$

Assuming that $\Delta(\vec{r}') = \Delta(x')$ and integrating over $y'$ and $z'$, we have

$$K_n = \frac{N_n}{4\beta}\sum_\omega \int_{-1}^{1} d\cos\theta \int_0^{2\pi} d\varphi \int \frac{d\vec{q}}{(2\pi)}\frac{\omega}{|\omega|(2i\omega+\vec{q}\vec{v}_n)}$$
$$\int_{-\infty}^{\infty} dx'\,\delta(q_z)\delta\left[eH_0(x+x') - q_y\right]e^{iq_x(x-x')}. \tag{A.2}$$

2. We shall integrate over $q_y$ and $q_z$. As a result, we obtain

$$K_n = \frac{N_n}{4}\frac{1}{\beta}\sum_\omega \frac{i\omega}{|\omega|}\int_{-1}^{1} d\cos\theta \int_0^{2\pi} d\varphi \int dx' \int dq_x\, e^{iq_x(x-x')}$$
$$\Delta(x')\,[2i\omega + v_n(q_x\cos\theta + eH_0(x+x')\sin\theta\cos\varphi)]^{-1}. \tag{A.3}$$

3. Let us integrate over $q_x$, passing to integrating in the complex plane and considering the cases *x-x' > 0* and *x-x' < 0*. Thereupon, we finally obtain

$$K_n = -\frac{N_n}{2v_n}\frac{1}{\beta}\sum_\omega \int_0^\pi \frac{\sin\theta\,d\theta}{\cos\theta} \int_0^{2\pi} d\varphi \int dx'\,\Delta(x')\,\frac{\omega}{|\omega|} \times$$

$$\times \left[\theta(x-x')\theta\left(\frac{-\omega}{\cos\theta}\right)e^{i(x-x')\left\{\frac{2i}{v_n}\left|\frac{\omega}{\cos\theta}\right|-eH_0(x+x')\,\mathrm{tg}\theta\cos\varphi\right\}} -\right.$$
$$\left. -\theta(x'-x)\theta\left(\frac{\omega}{\cos\theta}\right)e^{-i(x-x')\left\{\frac{2i}{v_n}\left|\frac{\omega}{\cos\theta}\right|+eH_0(x+x')\mathrm{tg}\theta\cos\varphi\right\}}\right]. \tag{A.4}$$

4. We shall sum over the Matsubara frequency ω = (2ν+1)πT, ν = 0,±1,± 2…Herein, we should consider ω>0 and ω<0 individually.
5. Thus, for example, at *x-x' > 0*, we have

$$\sum_\omega \frac{\omega}{|\omega|}\,\theta\left(-\frac{\omega}{\cos\theta}\right) e^{-\frac{2}{v_n}(x-x')\left|\frac{\omega}{\cos\theta}\right|} =$$

$$= \Sigma_{\omega>0}\left\{\theta\left(-\frac{\omega}{\cos\theta}\right) e^{-\frac{2}{v_n}(x-x')\left|\frac{\omega}{\cos\theta}\right|} - \theta\left(\frac{\omega}{\cos\theta}\right) e^{-\frac{2}{v_n}(x-x')\left|\frac{\omega}{\cos\theta}\right|}\right\} =$$

$$= [\theta(-\cos\theta) - \theta(\cos\theta)] \Sigma_{v>0} e^{-\frac{2}{v_n}\frac{(x-x')\pi T}{\beta|\cos\theta|} - \frac{4v\pi}{\beta|\cos\theta|}(x-x')} =$$

$$= \theta[-\cos\theta] - \theta[\cos\theta] \frac{e^{-\frac{2}{v_n}\frac{(x-x')\pi}{\beta v_n}}}{1-e^{-\frac{4}{v_n}\frac{(x-x')\pi}{\beta|\cos\theta|}}} = [\theta(-\cos\theta) - \theta(\cos\theta)] \frac{1}{2sh\frac{2(x-x')\pi}{\beta v_n|\cos\theta|}}. \quad (A.5)$$

Having summed in that way over ω in each term (A.4), we reduce this expression to the form

$$K_n = \frac{N_n}{4v_n} T \int_0^\pi \frac{\sin\theta\, d\theta}{\cos\theta} \int_0^{2\pi} d\varphi \int_{-\infty}^\infty \Delta(x')\, dx' \frac{1}{sh\frac{2|x-x'|\pi}{\beta v_n \cos\theta}} e^{-i(x^2-x'^2)eH_0 tg\theta \cos\varphi}. \quad (A.6)$$

The integrating over φ fields

$$K_n = \frac{N_n}{v_n\beta} \pi \int_1^\infty \frac{du}{u} \int_{-\infty}^\infty dx' \frac{\Delta_n^*(x') J_0[(x^2-x'^2)eH_0\sqrt{u^2-1}]}{sh\frac{2|x-x'|\pi u}{\beta v_n}} [\theta(x-x') - \delta_n] \quad (A.7)$$

Formulas (A.1) and (A.7) govern equation (2.7).

**Appendix B**

The analytical solutions to equation (3.21)

a) In the vicinity of the superconducting transition temperature ($T_c - T \ll T_c$), $\varepsilon\, \rho_n \ll 1$, $\rho_n \ll 1$ for the functions $f_{mn}$ (m; n = 1,2) on the basis of definitions (3.16)- (3.19), we obtain

$$f_{11}(\rho_1, \tilde{\varepsilon}) = \frac{35}{24}\varsigma(3)\tilde{\varepsilon}\rho_1 - \frac{109\cdot 31}{40\cdot 16}\varsigma(5)\tilde{\varepsilon}^2 \rho_1^2 + ...,$$

$$f_{12}(\rho_2) = \frac{7}{6}\varsigma(3)\rho_2 - \frac{31}{10}\varsigma(5)\rho_2^2 + ...,$$

$$f_{21}(\rho_1, \tilde{\varepsilon}) = \frac{7(3+2\tilde{\varepsilon})}{12(1+\tilde{\varepsilon})}\varsigma(3)\tilde{\varepsilon}\rho_1 - \frac{31(25+80\tilde{\varepsilon}+4\tilde{\varepsilon}^2)}{160(1+\tilde{\varepsilon})^2}\varsigma(5)\tilde{\varepsilon}^2 \rho_1^2 + ...,$$

$$f_{22}(\rho_2) = \frac{7}{6}\varsigma(3)\rho_2 - \frac{31}{10}\varsigma(5)\rho_2^2 + ..., \quad (B.1)$$

where $\varsigma(x)$ is the Riemann Zeta-function.

b) In the vicinity of the zero temperature $T \ll T_c$, $\tilde{\varepsilon}\, \rho_n \gg 1$, the asymptotes of the $f_{mn}$ functions have the form:

$$f_{11}(\rho_1\tilde{\varepsilon}) = \ln(1{,}12\sqrt{\tilde{\varepsilon}\rho_1\gamma}) - \frac{1}{\pi^2\rho_1\tilde{\varepsilon}}\left[\varsigma'(2) - \frac{\varsigma(2)}{2}\ln\left(\frac{\tilde{\varepsilon}\rho_1\gamma\pi^2}{2e_0^{\frac{1}{2}}}\right)\right] + ...,$$

$$f_{12}(\rho_2, \tilde{\varepsilon}) = \ln\left(\frac{2\sqrt{2\rho_2\gamma}}{e_0}\right) - \frac{1}{\pi^2\rho_2}\left[\varsigma'(2) - \frac{\varsigma(2)}{2}\ln\left(\frac{\rho_2\gamma\pi(1+\tilde{\varepsilon})}{4}\right)\right] + ...,$$

$$f_{21}(\rho_1, \tilde{\varepsilon}) = \ln(c(\tilde{\varepsilon})\sqrt{\tilde{\varepsilon}\rho_1\gamma}) - \frac{1+\tilde{\varepsilon}}{2\pi^2\rho_1\tilde{\varepsilon}^{3/2}}\left[\varsigma'(2) - \frac{\varsigma(2)}{2}\left(\ln\left(\frac{\tilde{\varepsilon}\rho_1\gamma\pi^2}{1+\tilde{\varepsilon}}\right) - \frac{\tilde{\varepsilon}}{2}\right)\right] + ...,$$

$$f_{22}(\rho_2) = \ln\left(\frac{2\sqrt{2\rho_2\gamma}}{e_0}\right) - \frac{1}{\pi^2\rho_2}\left[\varsigma'(2) - \frac{\varsigma(2)}{2}\ln\left(\frac{\rho_2\gamma\pi^2}{2}\right)\right] + ..., \quad (B.2)$$

where $c(\tilde{\varepsilon})=\dfrac{\sqrt{2.41(1+\sqrt{1+\tilde{\varepsilon}^2})}}{\sqrt{1+\tilde{\varepsilon}}}\exp(-0.205+\dfrac{1}{\tilde{\varepsilon}}\sqrt{1+\dfrac{1}{\tilde{\varepsilon}^2}})$, $e_0$ is the base of the natural logarithm, and $\ln\gamma = C$ is the Euler constant.

Let us pass to some simplifications assuming $\xi^{(1)}(T)=\xi^{(2)}(T)=\xi(T)$ and $\xi_c \to \xi^{(1)}(T_c)=\xi^{(2)}(T_c)=\xi_c$, then we shall consider each of the aforesaid temperature ranges individually. Case a) $\tilde{\varepsilon}\rho_n \ll 1, \rho_n \ll 1\, T_c - T \ll T_c$. In this case, the equation will take the form:

$$\lambda_{11}f_{11}(\rho_1,\tilde{\varepsilon}) + \lambda_{22}f_{22}(\rho_2) +$$
$$+ \xi_c(\lambda_{12}\lambda_{21}f_{21}(\rho_1,\tilde{\varepsilon}) - \lambda_{11}\lambda_{22}f_{11}(\rho_1,\tilde{\varepsilon})) + f_{22}(\rho_2)[\lambda_{22} - a\xi_c] = 0. \quad (B.3)$$

Substituting decompositions (B.1) into (B.3) and being restricted to the quadratic terms by the sought quantity $\rho_1 = \dfrac{v_1^2 e H_0}{(2\pi T)^2}$, we obtain the equation:

$$\tilde{\alpha}\rho_1^2 + \tilde{\beta}\rho_1 + \tilde{\chi} = 0, \quad (B.4)$$

where

$$\tilde{\alpha} = -\dfrac{31}{10}\varsigma(5)\left[\dfrac{\lambda_{22}-a\xi_c}{\lambda^4} - \dfrac{\tilde{\varepsilon}^2}{16}\xi_c\left(\dfrac{109}{4}\lambda_{11}\lambda_{22} - \dfrac{(25+80\tilde{\varepsilon}+4\tilde{\varepsilon}^2)}{(1+\tilde{\varepsilon})^2}\tilde{\lambda}_{12}\tilde{\lambda}_{21}\right) + \dfrac{109}{64}\tilde{\varepsilon}^2\lambda_{11}\right],$$

$$\tilde{\beta} = \dfrac{7}{6}\zeta(3)\left[\dfrac{\lambda_{22}-a\xi_c}{\lambda^2} - \dfrac{\tilde{\varepsilon}}{2}\xi_c\left(\dfrac{5}{2}\lambda_{11}\lambda_{22} - \dfrac{(3+2\tilde{\varepsilon})}{2(1+\tilde{\varepsilon})}\tilde{\lambda}_{12}\tilde{\lambda}_{21}\right) + \dfrac{5}{4}\tilde{\varepsilon}\lambda_{11}\right]$$

$$\tilde{\chi} = -v\left(\theta + \dfrac{\theta^2}{2}\right), \quad v = \sqrt{(\lambda_{11}-\lambda_{22})^2 + 4\tilde{\lambda}_{12}\tilde{\lambda}_{21}}. \quad (B.5)$$

The solution to equation (B.4) with subsequent expansion of this solution in terms of the quantity $\theta = 1-T/T_c$ gives us the behavior of the upper critical field $H_{c2}^{(ab)}$ (T) = $H_0$(T) in the vicinity of the superconducting transition temperature:

$$\rho_1 = \left(\dfrac{v\theta}{\tilde{\beta}} + \left(\dfrac{v(\tilde{\beta}^2 - 2\tilde{\alpha}v)}{2\tilde{\beta}^3}\right)\theta^2\right),$$

$$\rho_c^0 = \dfrac{v_1 v_2 e\, H_0(T)}{(2\pi T_c)^2} = \rho_1\dfrac{v_2}{v_1}\left(\dfrac{T}{T_c}\right)^2 = \left(\dfrac{v\theta}{\tilde{\beta}} + \left(\dfrac{v(\tilde{\beta}^2 - 2\tilde{\alpha}v)}{2\tilde{\beta}^2}\right)\theta^2\right)\lambda^{-1}\left(\dfrac{T}{T_c}\right)^2,$$

$$\lambda = {v_1}/{v_2}. \quad (B.6)$$

b) The range of low temperatures $T \ll T_c^0, \tilde{\varepsilon}\rho_n \gg 1$.

Substituting equation (B.2) into equation (3.21), it is not difficult to obtain the equation for determining the upper critical field $H_0(T) = H_{c2}(T)$ in the range of low temperatures

$$a(\ln x)^2 + B\ln x + \bar{C} + F(T) = 0, \quad (B.7)$$

where

$$a = \lambda_{11}\lambda_{22} - \tilde{\lambda}_{12}\tilde{\lambda}_{21}, B = 0{,}038\, a + v - \Lambda,$$
$$\bar{C} = \Lambda\,\xi_c - a\ln\lambda + \ln[\sqrt{\lambda}](0.04a - a\ln[\sqrt{\tilde{\varepsilon}}] + \lambda_{11} - \lambda_{22} + \Lambda) + 0.11\,\lambda_{11} -$$
$$- (0.4 - a\ln[\sqrt{\tilde{\varepsilon}}])(\Lambda + a\xi_c - \lambda_{22}),$$
$$\Lambda = \ln[c(\tilde{\varepsilon})]\,\tilde{\lambda}_{12}\tilde{\lambda}_{21} - 0{,}11\,\lambda_{11}\lambda_{22},\ x = \sqrt{\tilde{\varepsilon}\gamma\,\rho_c^0}\ .. \quad (B.8)$$

At the same time, F(T) is defined by the expression

$$F(T) = \tilde{\lambda}_{12}\tilde{\lambda}_{21}(\xi_c - [\ln x\sqrt{\lambda}c(\tilde{\varepsilon})] + q_{21})q_{12} - \left(\ln\left[\frac{2\sqrt{2x}}{e_0\sqrt{\lambda}\,\tilde{\varepsilon}}\right] - \xi_c\right)q_{21} -$$

$$-q_{22}\left((\xi_c - \ln[cx\sqrt{\lambda}])\lambda_{11} - 1\right)\lambda_{22} + q_{11}\lambda_{11}\left(\left(-\ln\left[\frac{2\sqrt{2x}}{e_0\sqrt{\lambda\tilde{\varepsilon}}}\right] + \xi_c + q_{22}\right)\lambda_{22} - 1\right), \quad (B.9)$$

where

$$x_0 = \sqrt{\tilde{\varepsilon}\gamma\,\rho_{c0}^0}, \quad \rho_{c0}^0 = v_1 v_2\, eH_0(0)/(2\pi T_c)^2,$$

$$q_{11}(T) = \frac{1}{\pi^2\rho_{c0}^0\tilde{\varepsilon}\lambda}\left(\frac{T}{T_c}\right)^2\left[\zeta'(2) - \frac{\zeta(2)}{2}\ln\left(\frac{\tilde{\varepsilon}\rho_{c0}^0\gamma\pi^2\lambda}{2e_0^{1/2}}\left(\frac{T_c}{T}\right)^2\right)\right],$$

$$q_{12}(T) = \frac{\lambda}{\pi^2\rho_{c0}^0}\left(\frac{T}{T_c}\right)^2\left[\zeta'(2) - \frac{\zeta(2)}{2}\ln\left(\frac{\rho_{c0}^0\gamma\pi^2(1+\tilde{\varepsilon})}{4\lambda}\left(\frac{T_c}{T}\right)^2\right)\right],$$

$$q_{21}(T) = \frac{(1+\tilde{\varepsilon})}{2\pi^2\rho_{c0}^0\tilde{\varepsilon}^{3/2}}\left(\frac{T}{T_c}\right)^2\left[\zeta'(2) - \frac{\zeta(2)}{2}\left(\ln\left(\frac{\tilde{\varepsilon}\rho_{c0}^0\gamma\pi^2\lambda}{1+\tilde{\varepsilon}}\left(\frac{T_c}{T}\right)^2\right) - \frac{\tilde{\varepsilon}}{2}\right)\right],$$

$$q_{22}(T) = \frac{\lambda}{\pi^2\rho_{c0}^0}\left(\frac{T}{T_c}\right)^2\left[\zeta'(2) - \frac{\zeta(2)}{2}\ln\left(\frac{\rho_{c0}^0\gamma\pi^2}{2\lambda}\left(\frac{T_c}{T}\right)^2\right)\right].$$

The solution to the equation for pure MgB$_2$ at T=0 is given by

$$\rho_{c0}^0 = \frac{v_1 v_2 eH_0(0)}{(2\pi T_c)^2} = (\gamma\tilde{\varepsilon})^{-1}\quad exp\left\{\frac{-B\pm\sqrt{B^2-4a\bar{c}}}{a}\right\} \quad (B.10)$$

At T<<T$_c$ for the upper critical field $H_{c2}^{(c)}(T) = H_0(T)$:

$$H_0(T) = H_0(0)\left[1 - \frac{2F(T)}{\sqrt{B^2-4a\bar{c}}}\right]. \quad (B.11)$$

**Figures**

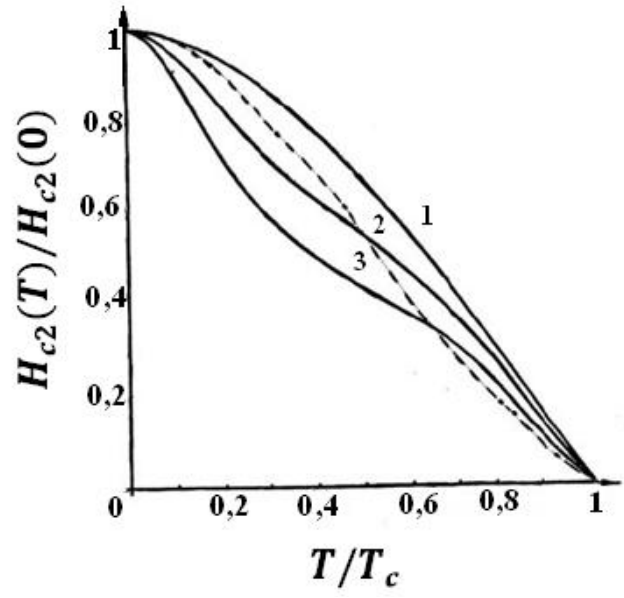

Fig.1

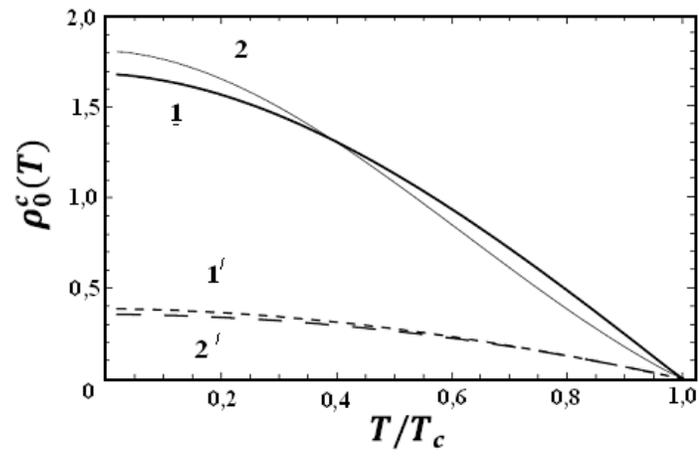

Fig.2

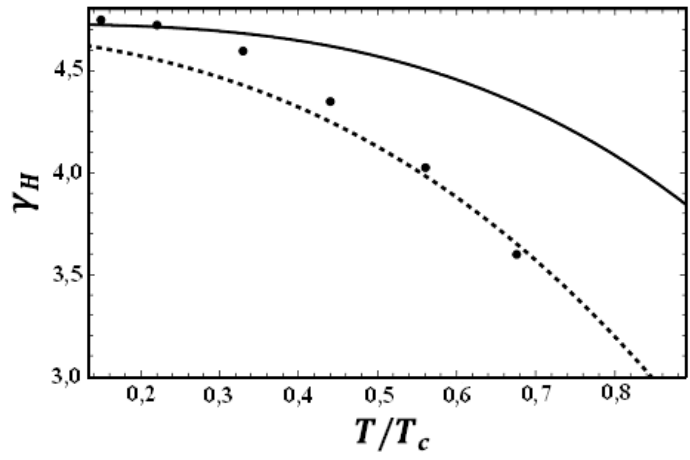

Fig.3.

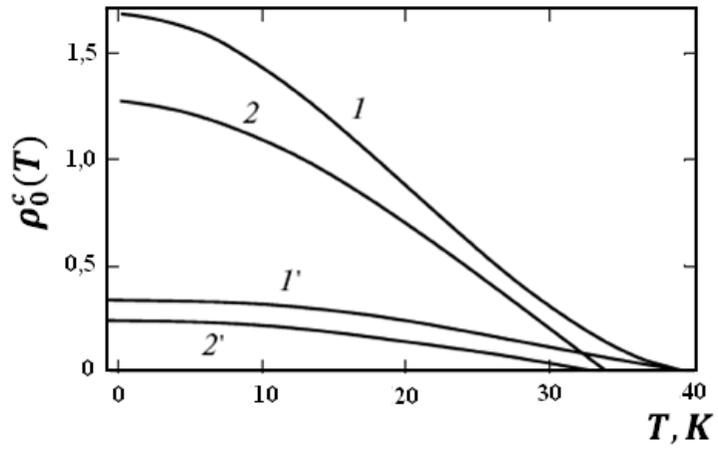

Fig.4

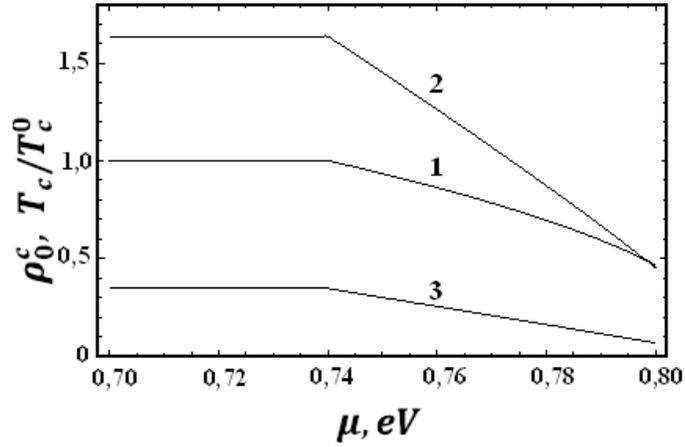

Fig.5

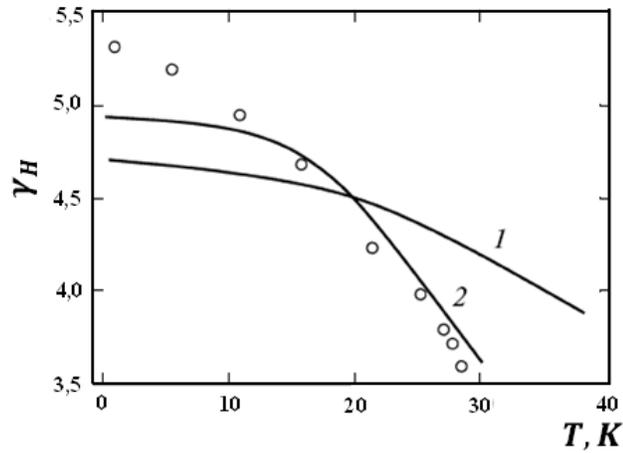

Fig.6

**Figure captions**

Fig.1. Dependence of the upper critical field $H_{c2}$ on temperature. The used parameters are $\lambda_{11} = 0.3$, $\lambda_{22} = 0$, $\lambda_{12} = 0.12$, $N_1/N_2 = 0.8$. The case $v_1/v_2 = 1$ (curve *1*)), $v_1/v_2 = 2$ (curve *2*), and $v_1/v_2 = 3$ (curve *3*). The dashed curve corresponds to the experimental dependence [39].

Fig.2. Dependence of the dimensionless upper critical field $\rho_c^0$ on temperature. Curves *1* and *2* correspond to the value $\rho_c^0$ (ab); curves *1'* and *2'* correspond to the value $\rho_c^0$ (c). Curves *1* and *1'* are plotted at the parameter value $\tilde{\varepsilon} = 0.31$ and the[ values of the renormalized interaction constants given as the a) set in the text (see Sect. 5); curves *2* and *2′* are plotted at the parameter value $\tilde{\varepsilon} = 0.133$ and the interaction constants given as the b)] set in the text (see Sect. 5).

Fig.3. Dependence of the anisotropy coefficient $\gamma_H = \rho_c(ab)/\rho_c(c)$ on temperature. The continuous curve is plotted at the values of the theory parameter $\varepsilon = 0.31$ and the value of the renormalized interaction constant given the *a]])* set in the text (see Sect. 5); the dashed curve is plotted at values of parameter $\tilde{\varepsilon} = 0.133$ and the parameters given as the *b)* set in the text (Sect. 5). The circles denote the experimental data [45].

Fig.4. Temperature dependences of upper critical fields $H_{c2}^{(ab)}$ and $H_{c2}^{(c)}$ (curves 1 and 1′, respectively) in pure MgB$_2$ ($\mu_0 = 0.74$ eV) and the same quantities (curves 2 and 2′) in doped MgB$_2$ ($\mu = 0.76$ eV). Parameters are the same as in Fig. 2.

Fig.5. Dependences of the ratio of the superconducting transition temperatures $\frac{T_C}{T_C^0}$ of doped and pure MgB$_2$ (curve 1) and critical fields $H_{c2}^{(ab)}$ (curve 2) and $H_{c2}^{(c)}$ (curve 3) at the temperature T = 0 on chemical potential $\mu$. Parameters are the same as in Fig. 2.

Fig.6. Temperature dependence of anisotropy coefficient for pure MgB$_2$ ($\mu_0 = 0.74$ eV, curve 1) and doped MgB$_2$ ($\mu = 0.76$ eV, curve 2). Parameters are the same as in Fig. 2. The circles correspond to experimental data borrowed from [43].